# $H_2^+$ embedded in a Debye plasma: Electronic and vibrational properties

M. L. Angel
and
H. E. Montgomery, Jr.[1]
Chemistry Program
Centre College
600 West Walnut Street
Danville, KY 40422 USA,

**Abstract**

The effect of plasma screening on the electronic and vibrational properties of the $H_2^+$ molecular ion was analyzed within the Born-Oppenheimer approximation. When a molecule is embedded in a plasma, the plasma screens the electrostatic interactions. This screening is accounted for in the Schrödinger equation by replacing the Coulomb potentials with Yukawa potentials that incorporate the Debye length as a screening parameter. Variational expansions in confocal elliptical coordinates were used to calculate energies of the $1s\sigma_g$ and the $2p\sigma_u$ states over a range of Debye lengths and bond distances. When the Debye length is comparable to the equilibrium bond distance, the plasma screening reshapes the potential energy curve. Expectation values, dipole polarizabilities and spectroscopic constants were calculated for the $1s\sigma_g$ state.

PACS numbers: 31.15.ae, 33.15.Dj, 52.40.Db

[1] Corresponding author: ed.montgomery@centre.edu
1(859)238-5731

## 1. Introduction

The interaction of atomic systems with plasmas has been studied extensively [1-9] and that work has recently been extended to molecular systems [10,11]. Plasma embedding is modeled by replacing the Coulomb potentials of the free system with Yukawa potentials [1,2] where the screening of the plasma is parameterized by the Debye length. It has been shown [10-14] that decreasing the Debye length is accompanied by a decrease in the binding energy. In this work we investigated the effect of plasma embedding on the potential energy curves of the $1s\sigma_g$ and $2p\sigma_u$ states of the $H_2^+$ molecular ion. $H_2^+$ was chosen because atomic and molecular hydrogen are abundant in interstellar matter and are of interest in astrochemistry[15]. The $1s\sigma_g$ state is the prototype for molecular bonding while the $2p\sigma_u$ state is an example of a van der Waals molecule [16] that is bound at internuclear separations greater than $R = 10.41\ a_0$. The effects of plasma embedding on the bonding behavior of these two states are described in this Letter. We also investigated the effects of Debye screening on the quadrupole moment integrals, parallel and perpendicular dipole polarizabilities, Dunham coefficients, harmonic force constants and harmonic frequencies of the $1s\sigma_g$ state.

## 2. Calculational method

Within the Born-Oppenheimer approximation, the Schrödinger equation in atomic units for $H_2^+$ is

$$\left\{-\frac{1}{2}\nabla^2 - \frac{1}{r_A} - \frac{1}{r_B} + \frac{1}{R}\right\}\psi = E\psi\ , \tag{1}$$

where $r_A$ and $r_B$ are the distances from the electron to nuclei $A$ and $B$ and $R$ is the internuclear separation. This equation is most easily solved in confocal elliptical coordinates defined as

$$\xi = \frac{r_A + r_B}{R}, \quad \eta = \frac{r_A - r_B}{R}, \quad 0 \le \phi \le 2\,\pi, \tag{2}$$

where $\phi$ is the angle of rotation about the internuclear axis. The Schrödinger equation for $\sigma$ states of $H_2^+$ is now

$$\left\{-\frac{2}{R^2\left(\xi^2-\eta^2\right)}\left[\frac{\partial}{\partial\xi}\left\{\xi^2-1\right\}\frac{\partial}{\partial\xi} + \frac{\partial}{\partial n}\left\{1-\eta^2\right\}\frac{\partial}{\partial\eta}\right] + \left[-\frac{4\,\xi}{R\left(\xi^2-\eta^2\right)} + \frac{1}{R}\right]\right\}\psi = E\psi\ . \tag{3}$$

Equation 3 is separable and has an analytic solution as the product of two infinite series [17,18].

When an atom or molecule is embedded in a hot, dense plasma, the electrostatic interactions between the particles are screened by the plasma. The Debye-Hückel theory – first formulated in the theory of electrolytes [19] – is widely used for modeling plasma screening because the required integrals are analytic. This Debye screening is included in the Schrödinger equation by replacing the Coulomb potentials by Yukawa potentials [1],

$$V(r) = \frac{e^{-\frac{r}{D}}}{r}, \tag{4}$$

where $D$, the Debye length, is a measure of the distance over which the Coulomb potential is killed off by the polarization of the plasma. Physically $D \propto \sqrt{T/n}$, where $T$ is the temperature of the plasma and $n$ is its number density. In the limit of large $D$ the Yukawa potential approaches the Coulomb potential and the plasma-embedded system approaches free $H_2^+$. However, if $D$ is comparable to the dimensions of the system, Debye screening significantly affects the interactions of the particles.

Incorporation of the Yukawa potential into the Schrödinger equation requires replacement of the potential energy portion of equation (3) by

$$V = -\frac{2}{R}\left[\frac{e^{-\frac{R(\xi+\eta)}{2D}}}{\xi+\eta} + \frac{e^{-\frac{R(\xi-\eta)}{2D}}}{\xi-\eta}\right] + \frac{e^{-\frac{R}{D}}}{R} \tag{5}$$

The resulting Schrödinger equation is non-separable and must be solved approximately.

The present work used a variational treatment with a trial wavefunction

$$\psi(\xi,\eta,\phi) = e^{-\alpha\xi}\sum_{\ell} c_{\ell}\; P_{\ell}(\eta)\left(\frac{\xi-1}{\xi+1}\right)^{m}, \tag{6}$$

where the $P_{\ell}(\eta)$ are the Legendre polynomials of degree $\ell$. This form of wavefunction was chosen so as to include terms similar to those used in the exact calculation [17,18]. For states of *gerade* parity, $\ell$ was restricted to even values while for states of *ungerade* parity, $\ell$ was odd. The $c_{\ell}$ were determined by solving the secular equation and $\alpha$ was hand optimized to give ten digit convergence at each value of $R$ and $D$. Using these wavefunctions to calculate the ground and excited state energies for unscreened $H_2^+$ gave variational energies that agreed with the exact energies [18] through $1\times10^{-9}\ E_h$.

For each value of $D$ we fitted a polynomial through the points around the minimum in the energy curves to find the equilibrium separation, $R_e$, and the energy at $R_e$.

The quadrupole moment integrals

$$\langle z^2 \rangle = \langle \psi | z^2 | \psi \rangle, \ \langle x^2 \rangle = \langle \psi | x^2 | \psi \rangle, \tag{7}$$

depend on the wavefunction at distances from the nuclei that are greater than those which effectively determine the energy [20] and their dependence on $D$ is thus of some interest. They can also be used in the Kirkwood approximation [21]

$$4\langle z^2 \rangle^2 \leq \alpha_{\parallel}, 4\langle x^2 \rangle^2 \leq \alpha_{\perp} \tag{8}$$

to obtain lower bounds to $\alpha_{\parallel}$ and $\alpha_{\perp}$, the parallel and perpendicular dipole polarizabilites.

The polarizabilities can be obtained from perturbation theory as

$$\alpha_{\parallel} = -2\langle \psi | \hat{z} | \psi^{\parallel} \rangle, \alpha_{\perp} = -2\langle \psi | \hat{x} | \psi^{\perp} \rangle, \tag{9}$$

where $\psi^{\parallel}$ and $\psi^{\perp}$ are the first-order corrections to $\psi$ under the perturbation operators $\hat{z}$ and $\hat{x}$, respectively. Hylleraas variational perturbation theory [22] has been shown [23] to be an accurate method to find $\psi^{\parallel}$ and $\psi^{\perp}$ and lower bounds to $\alpha_{\parallel}$ and $\alpha_{\perp}$ for $H_2^+$.

The Dunham parameterization [24] of the potential energy curve corresponds to a fourth-degree polynomial fit around the minimum in the potential energy where

$$E(R) = E(R_e) + A_0\rho^2\left(1 + A_1\rho + A_2\rho^2\right) \tag{10}$$

and

$$\rho = \frac{R - R_e}{R_e}. \tag{11}$$

The fitting parameters can then be used to find the harmonic force constant ($k_e$) and the harmonic frequency ($\omega_e$).

## 3. Results and discussion

$1s\sigma_g$ energies were calculated over the range $R = 1 - 10\ a_0$ for $D$ = 50, 20, 10, 5, 2 and 1. The calculation used the metric $\ell + m \leq 10$, $m$ = even, giving a 36-term wavefunction. $1s\sigma_g$ energies are given in Table 1 and are shown graphically in Figure 1. The effect of screening is more easily seen by calculating the potential energy curve $U(D,R)$ where

$$U(D,R)=E_{\mathrm{H}_2^+}(D,R)-E_{\mathrm{H}}(D), \quad (12)$$

and gives the energy of the screened molecular ion relative to $E_{\mathrm{H}}(D)$, the energy of a screened hydrogen atom with the same value of $D$ . The depth of these curves at $R_e$ is the dissociation energy of the embedded molecular ion, $D_e$. $E_{\mathrm{H}}(D)$ was calculated using the p-FEM method [25]. Results from the p-FEM method were consistent with [14] and with the recent work of Paul and Ho [26].

Potential energy curves for $D = \infty$, 5, 2 and 1 are shown in Figure 2. The curves for $D = 50$ and $D = 20$ were not shown as they are so close together at this scale as to be indistinguishable from the $D = \infty$ curve. For $D \geq 5\ a_0$, the effect of screening is small, with the change in binding energy of the molecular ion very nearly equal to the change in the energy of the Yukawa-screened hydrogen. For $D = 2$ and 1, the potential energy curves become markedly shallower and the minima in the curves shift to larger $R$. The changes in the potential energy curves result from an increase in the electronic energies (which include the kinetic energy, the electron-nuclear attraction and the energy of the screened hydrogen atom) offset by a slightly smaller decrease in the internuclear repulsion term, resulting in an overall decrease in stability. This decrease in stability with decreasing $D$ is consistent with the results observed by Mukherjee et al. [10].

$2p\sigma_u$ energies were calculated over the range $R = 10 – 20\ a_0$ for $D = 50$, 20, 10, 5, 4, 3 and 2. For $D = 2$, the excited state was non-bonding relative to the Yukawa-screened hydrogen atom. The metric was increased to 12 with $m$ = odd in order to get convergence at small $D$. This gave a 49-term wavefunction.

$2p\sigma_u$ energies are given in Table 2 and the potential energy curves for $D = \infty$, 10, 5, 4, 3 and 2 are shown in Figure 3. The $2p\sigma_u$ state is very loosely bound even at large $D$ and bonding occurs only at internuclear separations greater than 10 $a_0$. For $D > 10$. the dissociation energy initially increases and the equilibrium internuclear separation decreases as the Debye screening increases, For $D < 10$, the dissociation energy decreases accompanied by an increase in the internuclear separation.

As $D$ decreases in the region where $D > 10$, the internuclear repulsion term decreases more rapidly than the electronic term increases, resulting a small increase in stability. For $D < 10$, the $2p\sigma_u$ state behaves like the $1s\sigma_g$ state. In both cases, the changes in the dissociation energy are of the order of $10^{-4}$ $E_h$ while the equilibrium internuclear separation varies from 12.08 to 13.90 $a_0$.

Values of the equilibrium internuclear separation and dissociation energy are included in Tables 1 and 2.

Tables 3 – 5 show the dependence on $D$ of a variety of $1s\sigma_g$ electronic and spectroscopic properties. In all cases, they were evaluated at the value of $R_e$ corresponding to the equilibrium nuclear separation for that choice of $D$. Table 3 shows selected expectation values as a function of $D$. For $D \geq 5$ $a_0$, $\langle V \rangle$, $\langle z^2 \rangle$ and $\langle x^2 \rangle$ are relatively constant. However, when $D$ becomes comparable to the internuclear separation ($R \approx 2$ $a_0$), $\langle V \rangle$ drops significantly, while $\langle z^2 \rangle$ and $\langle x^2 \rangle$ show a marked increase. Since $\langle z^2 \rangle$ and $\langle x^2 \rangle$ measure the extent of the electron distribution and, through Eq. (8), provide lower bounds to the polarizability, we interpret this increase as consistent with a more diffuse, and thus more polarizable, electron distribution.

Table 4 shows the parallel and perpendicular dipole polarizabilities along with the Kirkwood lower bounds of Eq. (8). We have also shown the average polarizability

$$\alpha = \frac{1}{3}\left(\alpha_{\parallel} + 2\alpha_{\perp}\right), \tag{13}$$

and the anisotropy

$$\kappa = \frac{1}{3\alpha}\left(\alpha_{\parallel} - \alpha_{\perp}\right). \tag{14}$$

$\alpha_{\parallel}$, $\alpha_{\perp}$ and $\alpha$ show the same trends as $\langle z^2 \rangle$ and $\langle x^2 \rangle$, consistent with the conclusion of Kar and Ho [11] that the system becomes more polarizable as $D$ decreases. The anisotropy decreases with with decreasing $D$ and the electron distribution becomes more diffuse, approaching a near-spherical geometry for small $D$.

The Dunham parameters $A_0$, $A_1$ and $A_2$ are shown in Table 5 along with the harmonic force constant ($k_e = 2A_0/R_e^2$) and the harmonic frequency $\left(\omega_e = \sqrt{A_0/3.633\times10^{-16}}\right)$. Again we see a significant decrease in $k_e$ and $\omega_e$ at small $D$. Our free system values for $A_0 = 0.2054$ $E_h$, $A_1$ = - 1.7571 and $A_2 = 2.1264$ for $D = \infty$ are in good agreement with the values of [27] ($A_0 = 0.2053\ E_h$, $A_1 = -1.7363$ and $A_2 = 2.1329$).

In summary, the electronic and vibrational properties of $H_2^+$ embedded in a Debye plasma are significantly affected when $D \approx R_e$. The dissociation energy is reduced, the equilibrium internuclear separation increases, the polarizability increases and vibrational force constant is reduced. While the dipole polarizabilites calculated in this work were for a static electric field, it would be of interest to investigate effect of plasma embedding on the dynamic polarizabilies. This investigation is one of our goals.

**Acknowledgments**

H.E.M would like to thank K.D. Sen for his continued encouragement.

**References**

[1] J. C. Stewart, K.D. Pyatt, ApJ. 144 (1966) 1203.
[2] M.S. Murillo, J.C. Weisheit, Physics Reports 302 (1998) 1.
[3] P.K. Mukherjee, J. Karwowski, G.H.F. Diercksen, Chem. Phys. Lett. 363 (2002) 323.
[4] B. Saha, T.K. Mukherjee, P.K. Mukherjee, G.H.F. Diercksen, Theor. Chem. Acc. 108 (2002) 305.
[5] S. Ichimaru, Rev. Mod. Phys. 54 (1982) 1017.
[6] A.N. Sil, B. Saha, P.K. Mukherjee, Int. J. Quantum Chem. 104 (2005) 903.
[7] A.N. Sil, P.K. Mukherjee, Int. J. Quantum Chem. 106 (2006) 465.
[8] S. Kar, Y.K. Ho, Phys. Rev. A 77 (2008) 022713.
[9] S. Kar, Y.K. Ho, JQRST 109 (2008) 445.
[10] P.K. Mukherjee, S. Fritzsche, B. Fricke, Phys. Lett. A 360 (2006) 287.
[11] S. Kar, Y.K. Ho, Phys. Lett. A 368 (2007) 476.
[12] S. Skupsky, Phys. Rev. A 21 (1980) 1316.
[13]. K. Scheiber, J.C. Weisheit, N.F. Lane, Phys. Rev. A 35 (1987) 1252.
[14] M.A. Núñez, Phys. Rev. A 47 (1993) 3620.
[15] M. Padovani, D. Galli, A.E. Glassgold, Astronomy and Astrophysics 501 (2009) 619
[16] G. Herzberg, Spectra of Diatomic Molecules (D. Van Nostrand: New York, 1950) pp. 377-

381.

[17] D.R. Bates, K. Ledsham, A.K.Stewart, Phil. Trans. Roy. Soc. A246 (1953) 215.

[18] H. E. Montgomery, Chem. Phys. Lett. 50 (1977) 455.

[19] P. Debye, E. Hückel, Phys. Z. 24 (1923)185.

[20] A. Dalgarno, G. Poots, Proc. Phys. Soc. London A67 (1954) 343.

[21] J.G. Kirkwood, Phys. Z. 33 (1923) 57.

[22] E.A. Hylleraas, Z. Physik 65 (1930) 209.

[23] H.E. Montgomery, Chem. Phys. Lett. 56 (1978) 307.

[24] J.L. Dunham, Phys. Rev. 41 (1932) 721.

[25] M.N. Guimarães, F.V. Prudente, J. Phys. B: At. Mol. Opt. Phys. 38 (2005) 2811.

[26] S. Paul, Y.K. Ho, Phys. Plasmas 16 (2009) 063302.

[27] S. Mateos-Cortés, E. Ley-Koo, S. Cruz, Int. J. Quantum Chem. 86 (2002) 376.

**Table 1.** Energies in $E_{\text{h}}$ for $1s\sigma_g$ $H_2^+$ embedded in a Debye plasma

| $R$ ($a_0$) | $D = \infty$ | $D = 50$ | $D = 20$ | $D = 10$ | $D = 5$ | $D = 2$ | $D = 1$ |
|---|---|---|---|---|---|---|---|
| 1 | -0.45178631 | -0.43203847 | -0.40333621 | -0.35782414 | -0.27479473 | -0.07927245 | 0.10997427 |
| 2 | -0.60263421 | -0.58289414 | -0.55422709 | -0.50881491 | -0.42606393 | -0.23172997 | -0.04520628 |
| 3 | -0.57756286 | -0.55783451 | -0.52922719 | -0.48402329 | -0.40210546 | -0.21366909 | -0.04422103 |
| 4 | -0.54608488 | -0.52636777 | -0.49781893 | -0.45282386 | -0.37175069 | -0.18892452 | -0.03439361 |
| 5 | -0.52442030 | -0.50471245 | -0.47621058 | -0.43137647 | -0.35090744 | -0.17165594 | -0.02705181 |
| 6 | -0.51196905 | -0.49226768 | -0.46379661 | -0.41905766 | -0.33889097 | -0.16116181 | -0.02221179 |
| 7 | -0.50559400 | -0.48589640 | -0.45744133 | -0.41274289 | -0.33266080 | -0.15518356 | -0.01901731 |
| 8 | -0.50257039 | -0.48287454 | -0.45442560 | -0.40973585 | -0.32963609 | -0.15190162 | -0.01684549 |
| 9 | -0.50119545 | -0.48150009 | -0.45305208 | -0.40835735 | -0.32821205 | -0.15013570 | -0.01531628 |
| 10 | -0.50057873 | -0.48088349 | -0.45243369 | -0.40772991 | -0.32754236 | -0.14919346 | -0.01420328 |
| ∞ | -0.50000000 | -0.48029611 | -0.45181643 | -0.40705803 | -0.32680851 | -0.14811702 | -0.01028579 |
| | | | | | | | |
| $R_e$ | 1.996825 | 1.996933 | 1.997465 | 1.999247 | 2.006098 | 2.060332 | 2.360900 |
| $E(R_e)$ | -0.60263461 | -0.58289450 | -0.55422730 | -0.50881482 | -0.42606567 | -0.23188935 | -0.04831233 |
| $D_e$ | 0.10263461 | 0.10259839 | 0.10241087 | 0.10175678 | 0.09925716 | 0.08377233 | 0.03802654 |

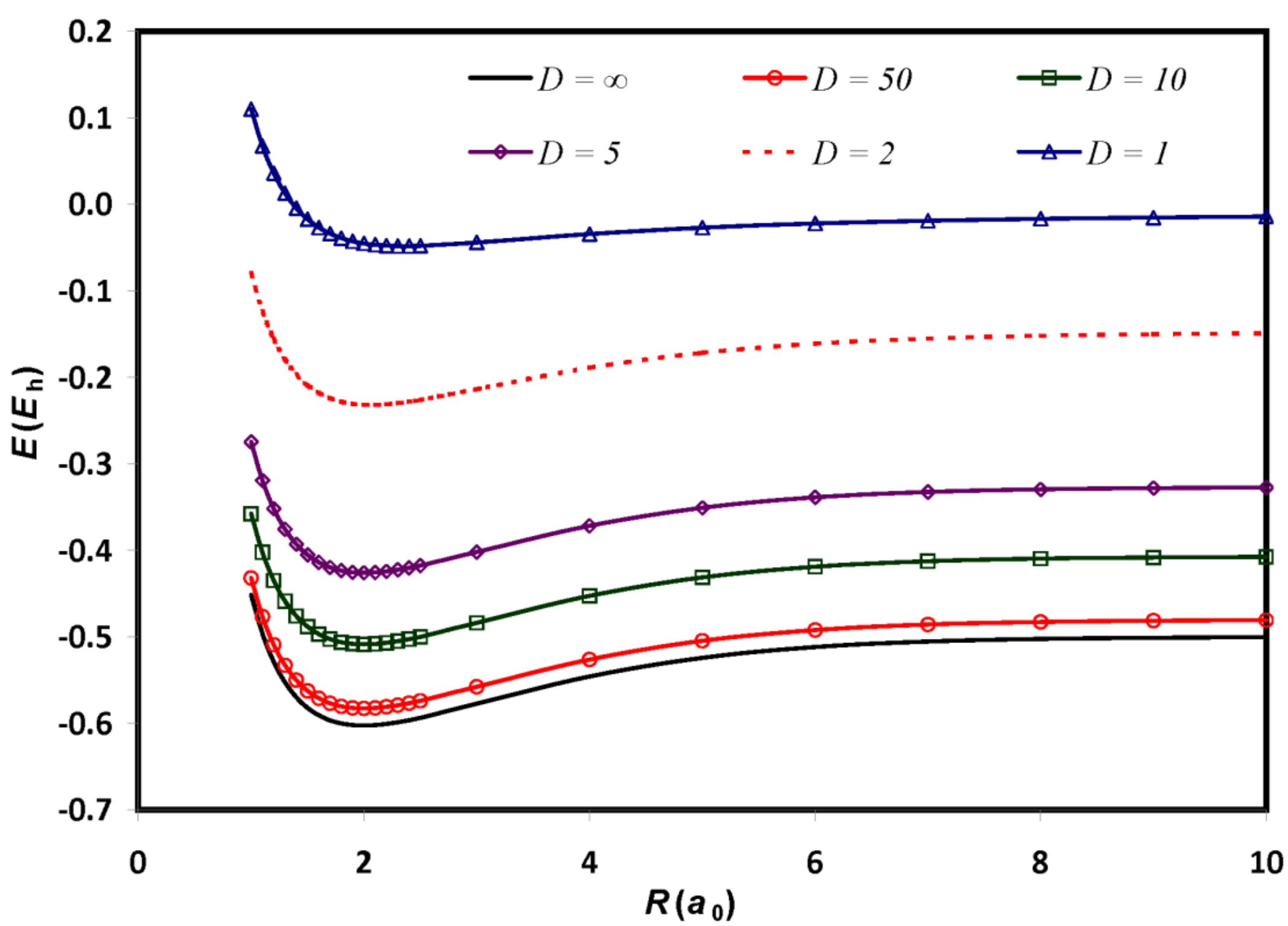

**Fig 1.** $1s\sigma_g$ energies

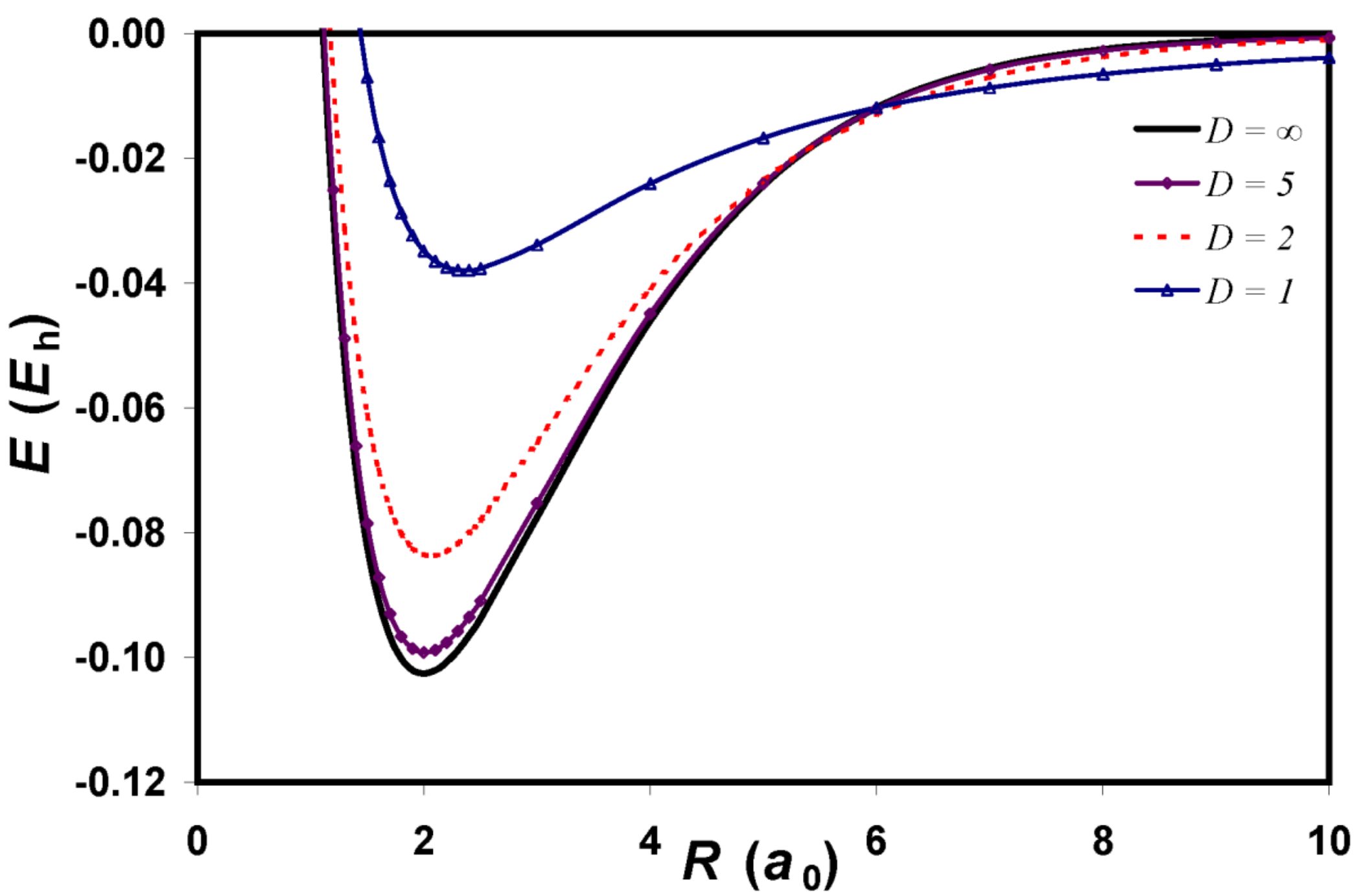

**Fig 2.** $1s\sigma_g$ $U(D,R)$ for $D = \infty$, 5, 2 and 1.

**Table 2.** Energies in $E_{\mathrm{h}}$ for $2p\,\sigma_u$ $H_2^+$ embedded in a Debye plasma

| $R$ ($a_0$) | $D = \infty$ | $D = 50$ | $D = 20$ | $D = 10$ | $D = 5$ | $D = 4$ | $D = 3$ | $D = 2$ |
|---|---|---|---|---|---|---|---|---|
| 10 | -0.49990107 | -0.48020631 | -0.45175695 | -0.40704313 | -0.32677122 | -0.29081730 | -0.23656384 | -0.14732128 |
| 11 | -0.50002442 | -0.48032909 | -0.45187710 | -0.40715783 | -0.32689160 | -0.29095505 | -0.23675433 | -0.14770032 |
| 12 | -0.50005789 | -0.48036205 | -0.45190768 | -0.40718307 | -0.32691837 | -0.29099315 | -0.23682727 | -0.14789715 |
| 13 | -0.50005947 | -0.48036316 | -0.45190655 | -0.40717685 | -0.32691160 | -0.29099390 | -0.23685097 | -0.14800024 |
| 14 | -0.50005150 | -0.48035475 | -0.45189606 | -0.40716168 | -0.32689521 | -0.29098278 | -0.23685521 | -0.14805475 |
| 15 | -0.50004206 | -0.48034488 | -0.45188426 | -0.40714571 | -0.32687813 | -0.29096965 | -0.23685261 | -0.14808364 |
| 16 | -0.50003368 | -0.48033606 | -0.45187369 | -0.40713147 | -0.32686322 | -0.29095788 | -0.23684826 | -0.14809905 |
| 17 | -0.50002688 | -0.48032881 | -0.45186483 | -0.40711946 | -0.32685102 | -0.29094825 | -0.23684401 | -0.14810730 |
| 18 | -0.50002155 | -0.48032295 | -0.45185750 | -0.40710944 | -0.32684124 | -0.29094063 | -0.23684034 | -0.14811172 |
| 19 | -0.50001742 | -0.48031813 | -0.45185135 | -0.40710099 | -0.32683336 | -0.29093456 | -0.23683717 | -0.14811407 |
| 20 | -0.50001420 | -0.48031396 | -0.45184597 | -0.40709367 | -0.32682679 | -0.29092949 | -0.23683423 | -0.14811528 |
| ∞ | -0.50000000 | -0.48029611 | -0.45181643 | -0.40705803 | -0.32680851 | -0.29091959 | -0.23683267 | -0.14811702 |
| $R_e$ ($a_0$) | 12.422056 | 12.396172 | 12.284674 | 12.081375 | 12.082440 | 12.392866 | 13.901264 | |
| $E(R_e)$ | -0.500061 | -0.480365 | -0.451910 | -0.407184 | -0.326919 | -0.290996 | -0.236855 | |
| $D_e$ ($R_e$) | -0.000061 | -0.000069 | -0.000093 | -0.000126 | -0.000111 | -0.000077 | -0.000023 | |

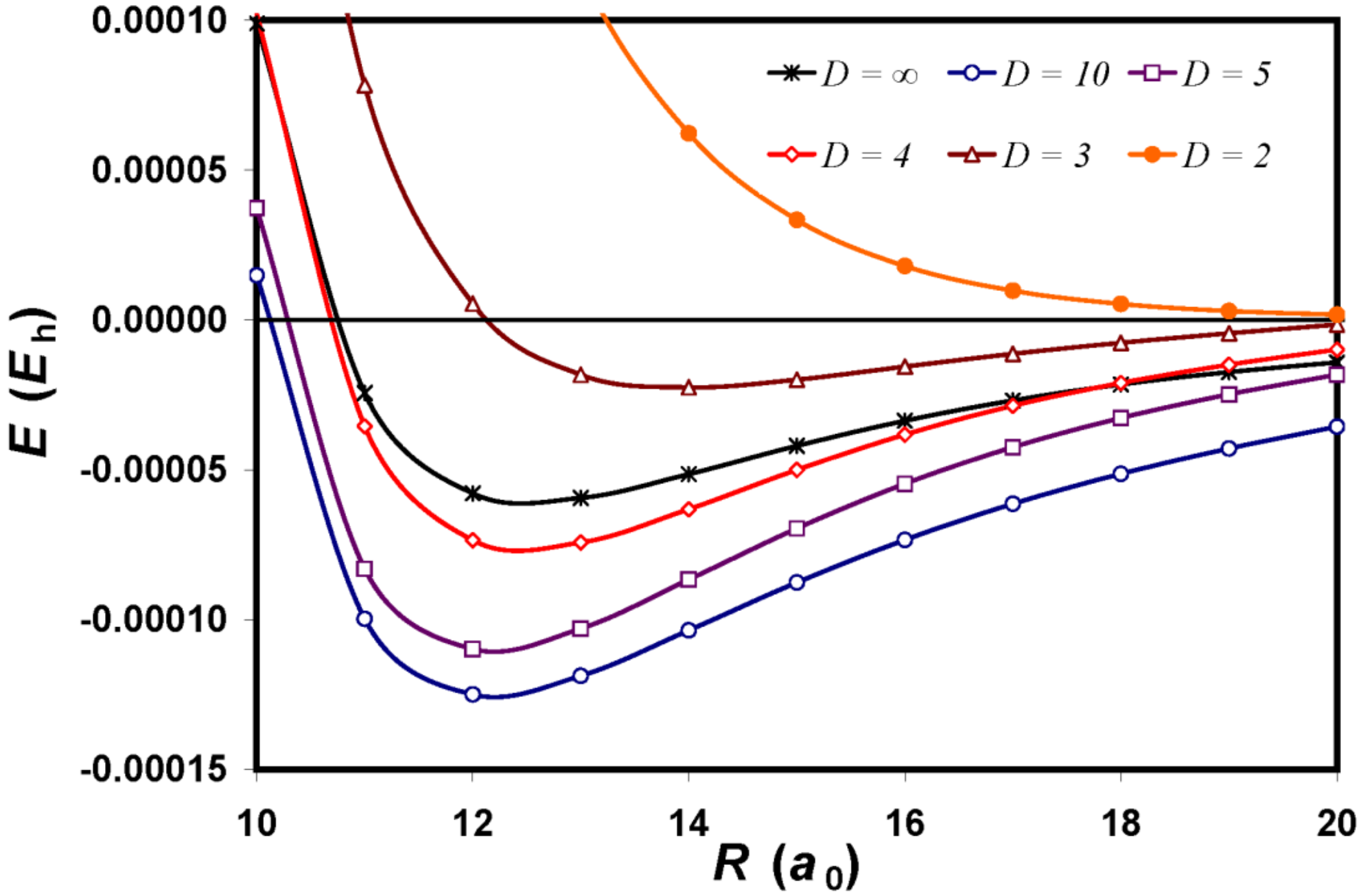

**Fig 3.** $2p\sigma_u$ $U(D,R)$ for $D = \infty$, 10, 5, 4, 3 and 2.

**Table 3.** 1s$\sigma_g$ expectation values for potential energy and quadrupole integrals

| $D$ | $\langle V \rangle$ | $\langle z^2 \rangle$ | $\langle x^2 \rangle$ |
|---|---|---|---|
| ∞ | -1.205268 | 1.108927 | 0.641143 |
| 50 | -1.185273 | 1.109463 | 0.641506 |
| 20 | -1.155327 | 1.112138 | 0.643330 |
| 10 | -1.105631 | 1.121110 | 0.649482 |
| 5 | -1.007629 | 1.153606 | 0.672235 |
| 2 | -0.728481 | 1.371570 | 0.831278 |
| 1 | -0.318519 | 2.804797 | 1.996692 |

**Table 4.** 1$s$ $\sigma_g$ dipole polarizabilities and Kirkwood lower bounds ($a_0^3$)

| $D$ | $\alpha_{\parallel l.b.}$ | $\alpha_{\parallel}$ | $\alpha_{\perp l.b.}$ | $\alpha_{\wedge}$ | $\alpha$ | $\kappa$ |
|---|---|---|---|---|---|---|
| ∞ | 4.918876 | 5.057482 | 1.644256 | 1.754475 | 2.855477 | 0.385576 |
| 50 | 4.923633 | 5.062401 | 1.646120 | 1.756608 | 2.858539 | 0.385487 |
| 20 | 4.947400 | 5.086966 | 1.655496 | 1.767330 | 2.873875 | 0.385036 |
| 10 | 5.027550 | 5.169825 | 1.687306 | 1.803680 | 2.925728 | 0.383511 |
| 5 | 5.323223 | 5.475791 | 1.807601 | 1.941421 | 3.119544 | 0.377659 |
| 2 | 7.524813 | 7.776257 | 2.764093 | 3.060827 | 4.632637 | 0.339291 |
| 1 | 31.467546 | 35.103927 | 15.947112 | 20.734794 | 25.524505 | 0.187651 |

**Table 5.** 1$s$ $\sigma_g$ Dunham parameters, harmonic force constant ($k_e$) and harmonic frequency ($\omega_e$)

| $D$ | $A_0$ ($E_h$) | $A_1$ | $A_2$ | $k_e$ ($E_h/a_0^2$) | $\omega_e$ (cm$^{-1}$) |
|---|---|---|---|---|---|
| ∞ | 0.2054 | -1.7571 | 2.1264 | 0.1030 | 2325 |
| 50 | 0.2054 | 1.7573 | 2.1273 | 0.1030 | 2324 |
| 20 | 0.2052 | -1.7580 | 2.1316 | 0.1028 | 2322 |
| 10 | 0.2044 | -1.7608 | 2.1467 | 0.1022 | 2316 |
| 5 | 0.2012 | -1.7725 | 2.2085 | 0.1000 | 2290 |
| 2 | 0.1761 | -1.8435 | 2.7753 | 0.0831 | 2088 |
| 1 | 0.0936 | -2.1862 | 3.7030 | 0.0338 | 1332 |